\documentclass[aps,prb,showpacs,twocolumn]{revtex4-1}
\usepackage{amssymb}
\usepackage{amsmath}
\usepackage{graphicx}
\usepackage{dsfont}
\usepackage{times}

\begin{document}

\title{Extracting Critical Exponent by Finite-Size Scaling with
Convolutional Neural Networks}
\author{Zhenyu Li,$^1$ Mingxing Luo,$^1$ and Xin Wan$^{1,2}$}
\affiliation{$^1$Zhejiang Institute of Modern Physics, Zhejiang University,
Hangzhou 310027, China}
\affiliation{$^2$Collaborative Innovation Center of Advanced Microstructures,
  Nanjing University, Nanjing 210093, China}
\date{\today }

\begin{abstract}
Machine learning has been successfully applied to identify phases and phase
transitions in condensed matter systems.
However, quantitative characterization of the critical fluctuations near phase
transitions is lacking.
In this study we propose a finite-size scaling approach based on a
convolutional neural network and analyze the critical behavior of a quantum
Hall plateau transition.
The localization length critical exponent learned by the neural network is
consistent with the value obtained by conventional approaches.
We show that the general-purposed method can be used to extract critical exponents
in models with drastically different physics and input data,
such as the two-dimensional Ising model and 4-state Potts model.
\end{abstract}

\maketitle

\section{Introduction}
\label{sec:intro}
Recent studies have established machine learning as an effective tool
to identify phases and phase transitions.
The prototypical two-dimensional Ising model has been studied, e.g., by
the restricted Boltzmann machine (RBM),~\cite{Torlai16,Morningstar17}
the feed-forward sigmoid neural network,\cite{Nieuwenburg17}
the convolutional neural network (CNN),~\cite{Carrasquilla17,Tanaka17}
the principal component analysis,~\cite{Wang16}
and  the supporting vector machine.~\cite{Ponte17}
Machine learning methods have also been applied to study
the Potts model,~\cite{Li17}
the Ashkin-Teller model,~\cite{Rao17}
the transverse-field Ising model,~\cite{Carleo17,Schmitt17}
the Kitaev toric codes,~\cite{Deng16,YiZhang17}
the Hubbard model,~\cite{broecker17,Chng17,Saito17,Nomura17}
and disordered electron systems~\cite{Ohtsuki16,Ohtsuki17,Yoshioka17,Mano17}
among others.
The wide success in the identification of phases and
the location of the phase transitions is possibly rooted
in ideas like renormalization group.~\cite{beny13,mehta14,lin16}
Connections of neural network states and tensor network states
have also been explored.~\cite{Chen17,Gao17,Huang17,Deng17}

One of the advantages of machine learning is that
one can provide raw low-level data,
such as spin configurations, energy spectrum, or wave functions,
so that only elementary knowledge in physics is required.
With sufficiently large data sets, higher-level features can be recognized by
various deep learning architectures~\cite{Goodfellow}
and then be used to distinguish phases.
Nevertheless, quantitative understanding of
the critical behavior and, hence, the identification of universality classes
have been scarce in the past studies.

In disordered electronic systems,
conventional approaches calculate the Lyapunov exponent,~\cite{kramer93}
the inverse participation ratio (IPR),~\cite{wegner80}
the Thouless number,~\cite{licciardello75}
and the Chern number.~\cite{huo92}
Interestingly, the IPR, which measures the occupation of
a wave function in real space, can be cast into a
neural network representation with low-level input like eigenstates.
Fig.~\ref{fig1}(a) illustrated that the IPR of a normalized eigenstate
$\vert \psi \rangle = \sum_i c_i \vert i \rangle$, defined as
${\rm IPR} = \sum_i \vert c_i \vert^4$,
can be mapped to a single artificial neuron with
the square of the local electron density $\vert c_i \vert^4$ as input.
The weights for the inputs are all unity and a step-function activation
can be used to {\em semi-quantitatively} distinguish the conducting states from
the localized states. One expects that for a $d$-dimensional lattice
with linear size $L$, the IPR of a conducting state tends to $L^{-d}$,
while that of a localized state is a finite constant.

At the Anderson transition, there are strong fluctuations in the local density
of states.
The simple concept of the IPR can be generalized to
the whole multifractality spectrum
at the critical point to characterize infinitely many
relevant operators.~\cite{Jensen94,huckestein95,evers08}
Such a generalization developed by insightful physicists is
not expected to be learned by a machine simply from large-scale data sets.
But with cleverly designed algorithms, can machine display intelligence
in the study of critical exponent?
Close to a critical energy $E_c$, the localization length $\xi$,
or the extent of wave functions, diverges as $\xi(E) \propto
\vert E - E_c \vert^{-\nu}$.~\cite{huckestein95}
The cutoff of the singular behavior by the finite system size $L$
in computation leads to signatures in physical observables.
One expects that the width of the critical regime defined by
$\xi(\Delta E) = L$ is also encoded in some machine-learning
observables.

In this paper, we combine the conventional wisdom and the
novel machine learning techniques to study the critical behavior of
the quantum Hall plateau transition
in a two-dimensional disordered electron system.
The approach can be thought of as a generalization of the IPR study
based on a CNN architecture.
The learning is guided by a trial-and-error labelling scheme,
which detects the cutoff in the length scale by finite system size.
A localization length critical exponent $\nu = 2.22 \pm 0.04$ can be extracted
from the finite-size scaling of the characteristic energy scale where the
localization length is comparable to the system size.
We explain how machine detects the critical fluctuations and identifies the
relevant scales, and show that the same scheme can be applied to
other models with drastically different physics.
In particular, using spin configurations as input,
we obtain the correlation length exponent
$\nu = 0.985 \pm 0.002$ for the Ising model on a square lattice and
$\nu = 0.69 \pm 0.01$ for the 2D 4-state Potts model.

The paper is organized as follows.
We present the lattice model for the plateau transition,
the neural network architecture, and the trial-and-error learning scheme
in Sec.~\ref{sec:model}.
In Section~\ref{sec:results}, we show the training results and
explain how to read the length-scale information,
from which we can extract the localization length critical exponent.
We summarize the results and discuss the general applicability of the
critical exponent extracting method in Sec.~\ref{sec:conclusion}.
In Appendix~\ref{app:network}, we present the technical details
of the architecture, dataset preparation, training, and weight initialization
for the CNN, as well as the random errors in the location of the
worst performance.
We simulate the performance curve in a phenomenological analysis
in Appendix~\ref{app:modeling}.
In Appendix~\ref{app:robust}, we demonstrate the robustness of our results
against network architecture,
activation function, and filter size and number of channels.
Finally, we discuss the application of the same neural network architecture
and the same trial-and-error method to extract the correlation length
critical exponent for the two-dimensional (2D) Ising model and
4-state Potts model in Appendix~\ref{app:spin}.

\section{Model and Method}
\label{sec:model}

We consider the disordered Hofstadter model on a square lattice with Hamiltonian
\begin{equation}
H_0 = - \sum_{\langle i,j\rangle}\{e^{i\theta_{ij}}c_i^\dagger c_j+h.c.\}
+ \sum_i\epsilon_i c_i^\dagger c_i,
\label{eq:Hofstadter_model}
\end{equation}
where the quantized flux per plaquette is $1/3$.
We project the on-site disorder to the lowest of the three magnetic subbands
and suppress the kinetic energy such that the corresponding pure system
is a Chern insulator with Chern number 1 and a perfectly flat band.
The on-site random potential $\epsilon_i$ is distributed uniformly
in [-0.5, 0.5].
In this model, extended states only exist at the band center $E_c = 0$.
When the Fermi energy sweeps across the band center, the system exhibits a
localization-delocalization-localization transition.
The detail of the model and its critical behavior obtained by Thouless number
calculation are available in Ref.~\onlinecite{bhatt02}.
For each system size, we diagonalize the Hamiltonian with sufficiently many
independent realizations of disorder potential and
randomly select 335,000 normalized wave functions:
300,000 for training and 35,000 for testing.
We record the corresponding energy for each state so we can generate
the density of conducting or localized states later.
Motivated by the neuron representation of the IPR, we feed the square of the
electron density on each site to a multilayer CNN, which is capable of
extracting more complex features than the IPR.

We construct the CNN with two convolutional layers.
The first layer convolves 16 $3 \times 3$ filters (with stride 1),
followed by rectified linear unit (ReLU) activation.
The second layer convolves 18 $3 \times 3$ filters (with stride 1), also
followed by ReLU activation.
The outcome is then flattened into a fully connected layer,
followed by another fully connected layer of 256 nodes with ReLU activation,
which is fully connected to a logistic output layer.
The detail of the CNN is presented in Appendix~\ref{app:network}.
The loss function is defined by the cross entropy of the output
and the input labels of conducting or localized states, supplemented by the
L2-regularization of the weights between the fully connected layers.
We implement the CNN with TensorFlow and use the Adam algorithm
to optimize the loss for the training set.
We denote the CNN output, before the Softmax function,
as $y_1$ and $y_2$, whose values are indicators of localized and conducting
states, respectively.
We define the performance of the CNN by the rate of correct identification
on the test set.

\begin{figure}
\centering
\includegraphics[width=8cm]{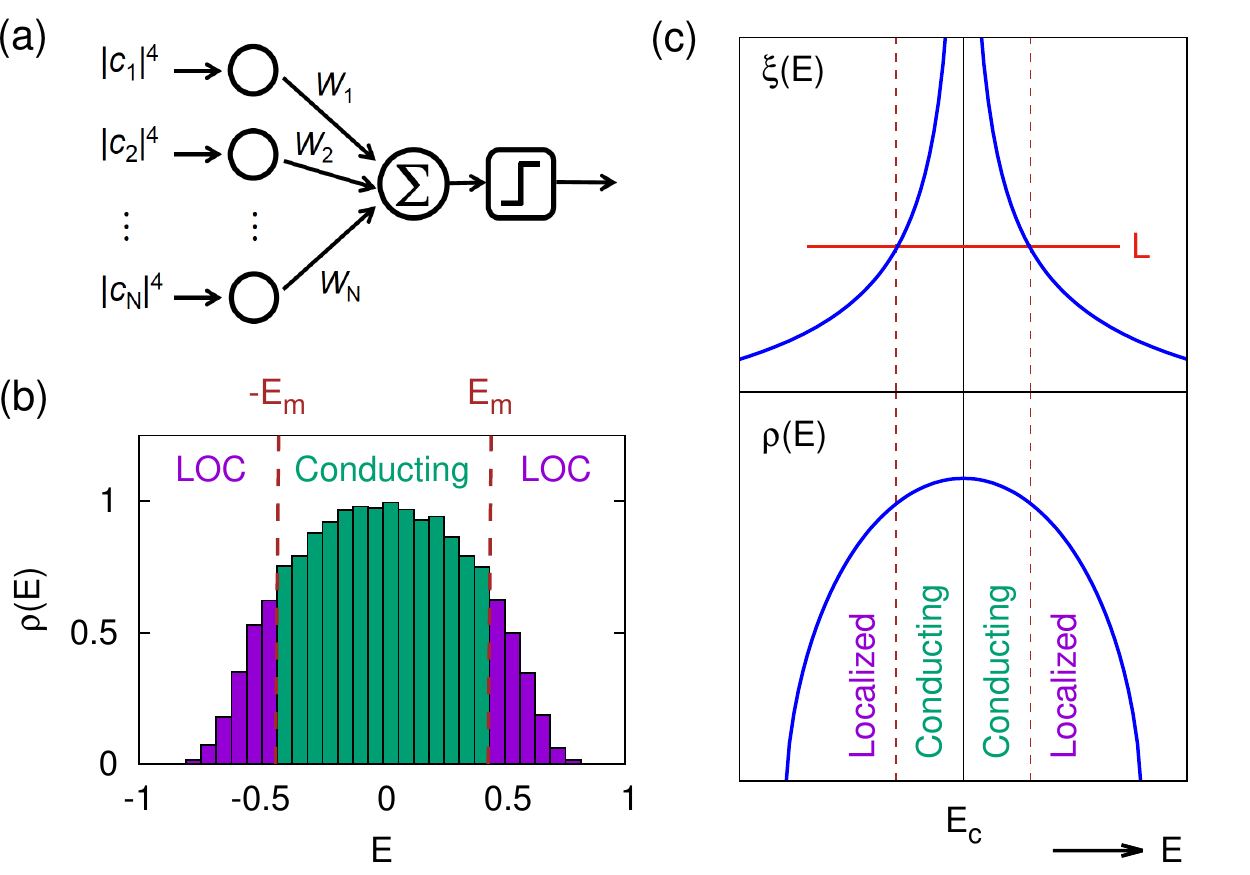} %
\caption{
(Color online.)
Illustration of the perceptron model for calculating IPR and
the scheme to detect the characteristic energy scale
and its physics motivation.
(a) The IPR can be expressed as a perceptron whose inputs are
the square of the local electron density and
whose weights $W_i$s are all set to be unity.
A step-function activation can be introduced to roughly
distinguish the localized states from the conducting ones.
(b) We input labels for the eigenstates within $-E_m < E < E_m$ as conducting
states and vary $E_m$ to observe the performance of the CNN.
(c) The labelling scheme is motivated by the fact that
the localization length $\xi \sim \vert E - E_c \vert^{-\nu}$
as energy $E$ approaches the critical point $E_c$.
The states with $\xi$ greater than the system size $L$ are conducting;
otherwise, they are localized.
}
\label{fig1}
\end{figure}

One way to extract critical exponent numerically is to characterize
fluctuations in finite systems and to perform finite-size scaling.
For example, one can calculate topological Chern numbers to identify
eigenstates in a quantum Hall system as being conducting or localized
and obtain the localization length critical exponent by the finite-size
scaling of the characteristic width of the density of the conducting
states.~\cite{huo92}
Without such a Chern number or conductance calculation,
we do not {\it a priori} know which states are localized
and which states are conducting.
To pursue a supervised learning, we need to label the states;
this requires the knowledge of the conducting or the localized
nature of the wave functions.
However, explicitly acquiring such information by conventional
approaches and supplying it to the CNN would defy
the necessity of machine learning in the first place.
To circumvent the difficulty,
we introduce a trial-and-error method to identify the characteristic
width for conducting states.
We label the eigenstates within $-E_m < E < E_m$ as conducting
states, as illustrated in Fig.~\ref{fig1}(b),
and vary $E_m$ to study the CNN performance,
according to the postulated labels.
The labelling scheme is motivated by the sketch
in Fig.~\ref{fig1}(c).
In the vicinity of the critical energy $E_c$, the localization length
$\xi$ diverges with critical exponent $\nu$.
The finite system size $L$ provides a cutoff to the length scale.
Close to $E_c$, when $\xi$ is greater than $L$, the eigenstates
are conducting.
Otherwise, they are localized.
We will show that the resulting performance curve can be used to
extract the characteristic energy scale and hence $\nu$.

\section{Results}
\label{sec:results}

\begin{figure}
\centering
\includegraphics[width=8cm]{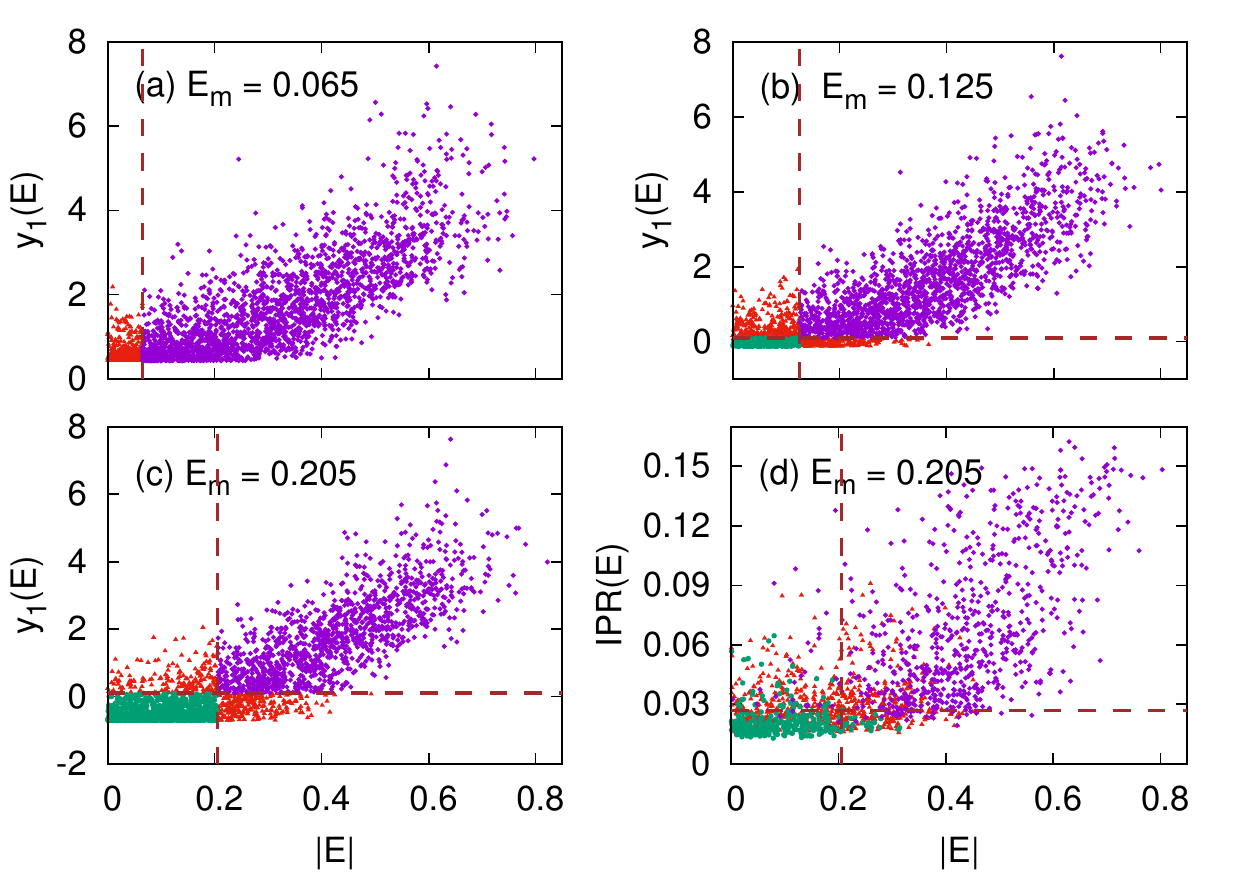} %
\caption{
(Color online.)
CNN output $y_1$ of 2,500 test states as a function of energy $E$
for (a) $E_m = 0.065$, (b) $E_m = 0.125$, and (c) $E_m = 0.205$.
The color scheme is such that the CNN correctly identifies the green points as
conducting states and purple points as localized states,
but misidentifies the red points.
The horizontal dashed lines in (b) and (c) indicate $y_1 = y_2$.
(d) The scatter plot of the IPR for $E_m = 0.205$.
The color of a point is green if the state is classified as conducting state
with over 80\% probability, purple as localized state with over 80\%
probability, or red if neither is true.
The horizontal line is guide to eye
as the separation of the conducting and localized states
according to the CNN output.
In all panels, the vertical lines mark $E_m$.
}
\label{fig2}
\end{figure}

\begin{figure}
\centering
\includegraphics[width=8.0cm]{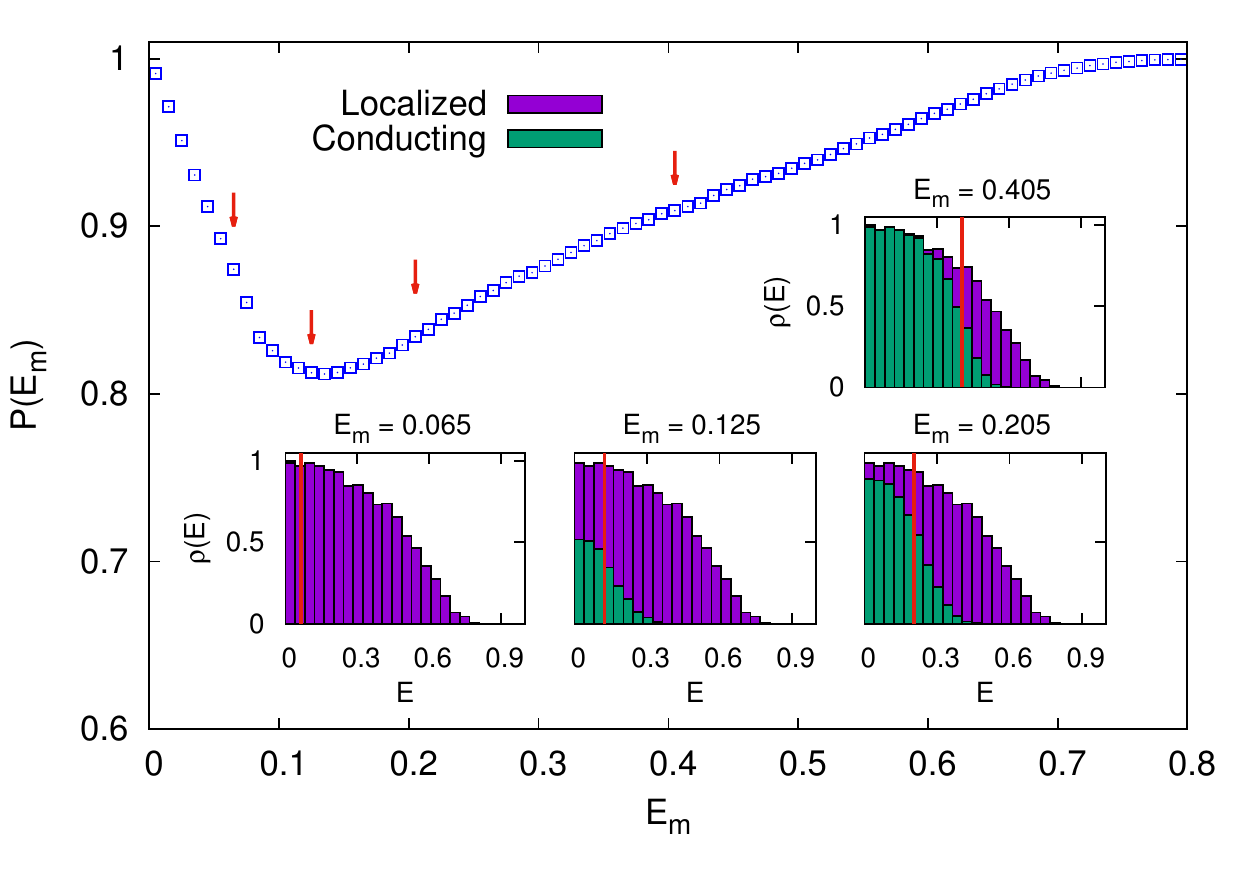} %
\caption{
(Color online.)
The characteristic V-shape of the performance curve $P(E_m)$.
As $E_m$ increases, conducting states identified by the CNN emerge and,
eventually, occupy the whole band,
as indicated by the varying density of conducting (localized) states
in the insets.
}
\label{fig3}
\end{figure}

We start with a $12 \times 12$ lattice and diagonalize the system with
different disorder realizations and randomly
select 300,000 wave functions for training.
We then feed the square of the local electron density to the CNN and
obtained the optimal CNN parameters and plot the CNN output $y_1(E)$
for 2,500 eigenstates as scattered points in Fig.~\ref{fig2} for
selective $E_m$.
The spread of the data points is similar for various $E_m$ but
their values, especially near the lower bounds, change.
The states are identified by CNN as localized states if $y_1 > y_2$,
or otherwise conducting; the density of the localized and conducting states
are shown in the insets of Fig.~\ref{fig3}.
At $E_m = 0.065$, all states are identified as localized states,
even though only those with $\vert E \vert > E_m$ [purple diamonds in
Fig.~\ref{fig2}(a)]
are labelled as localized states in the input,
whose percentage represents the performance of the CNN.

As $E_m$ increases, the CNN eventually realizes that there are
conducting states near the band center.
However, there is a fundamental conflict.
As shown in Fig.~\ref{fig2}(b) and (c), the localized and
conducting states are separated by a vertical line $|E| = E_m$
according to the input labels.
However, they are distinguished by a horizontal line
(indicating $y_1 = y_2$~\cite{y1y2}) according to the CNN output.
Therefore, the two red areas in Fig.~\ref{fig2}(b) and (c) contain the states
that are labelled as conducting but classified as localized, or vice versa.

The physics behind the conflict in the CNN identification is nothing but
the ubiquitous fluctuations, also present in conventional quantities,
such as the IPR.
Figure~\ref{fig2}(d) plots the IPR with the following coloring scheme:
The states with more than 80\% probability of being conducting (localized) are
represented by green dots (purple diamonds); the rest by red triangles
are states with less certain classifications.
The probabilities are given by the logistic outputs of the CNN with
$E_m = 0.205$, represented by the vertical line in Fig.~\ref{fig2}(d).
We also include a horizontal line IPR $= 0.027$ (for illustration)
to further separate the plot area into four regions.
The green (purple) points, for which the CNN output agrees with the input
labels, are distributed mainly in the lower left (upper right) regions.
The red points, for which fluctuations lead to contradiction
between the CNN output and the labels,
appear in all four regions in similar numbers.
In agreement with Fig.~\ref{fig2}(c), they dominate in
the upper left and lower right regions, but are outnumbered
in the other two regions.
The comparison shows that the CNN output $y_1$ (or $y_2$)
can be thought of as a generalization of the IPR,
but with a clear advantage that the spread of $y_1$ is greatly reduced
from that of the IPR.

We test a set of 35,000 wave functions, independent of the training set,
to obtain the performance $P(E_m)$ of the CNN, as shown in Fig.~\ref{fig3}.
The density of the conducting/localized states, according to the CNN
predictions, is shown in the insets for four selective $E_m$.
Unlike the W-shaped curve in Ref.~\onlinecite{Nieuwenburg17},
$P(E_m)$ has an asymmetric V shape.
We note that the scheme we use resembles the
confusion scheme in Ref.~\onlinecite{Nieuwenburg17}, but with a
significant difference.
The confusion scheme was introduced to study the transition between
two distinct phases.
Here, there is essentially only one phase, i.e. the localized phase
(extended states exist only at $E = 0$),
and we are studying the fluctuations in finite systems.
The minimum of the V-shaped curve defines the only energy scale
$E_m^{\rm min} \approx 0.125$ for a given system size.
Therefore, one would expect that $E_m^{\rm min}$ should
correspond to $\xi \sim L$ for an effective learning,
such that states are conducting when $E < E_m^{\rm min}$
and localized when $E > E_m^{\rm min}$, in perfect agreement with the
input labels.
But the paradox is that such a seemingly perfect labelling results
in the worst performance. Why?

The key in the answer is that the CNN is not provided with energies,
which determine labels, and, therefore, can only classify
according to the features that it learns from wave functions.
According to the analysis of Fig.~\ref{fig2}, the performance of the CNN is
a compromise between the two possible classification scenarios.
For small $E_m$, the CNN concludes that all states are localized
due to the overwhelming localized states in the training data,
hence $P(E_m)$ decreases with increasing $E_m$.
We can estimate the performance $P(E_m)$ for small $E_m < E_m^{\rm min}$ by
\begin{equation}
P_{<}(E_m) = 1 - 2 E_m \rho(E_m),
\end{equation}
because the density of states $\rho(E)$ is flat near the band
center [see Fig.~\ref{fig1}(b)].
For large $E_m$, the deficiency in learning is caused by the fluctuations,
such that states with the same energy can yield different CNN output,
or states with the same CNN output can spread in energy.
If we denote the performance due to this mechanism as $P_{>}(E_m)$,
we have approximately
\begin{equation}
1 - P_{>}(E_m) \sim \rho(E) \delta y_1 \left [{{d \langle y_1 \rangle} \over dE} \right ]^{-1}
\end{equation}
at $E = E_m$, which depends on
the energy dependence of the mean value of $y_1$,
and the amplitude of its fluctuations $\delta y_1$ [see Fig.~\ref{fig3app}
in Appendix~\ref{app:modeling}].
As we approach $E_c = 0$, $\xi$ increases as $\vert E \vert^{-\nu}$
and exceeds $L$ at $\vert E \vert  = E_L \sim L^{-1/\nu}$.
Meanwhile, we expect the mean value of $y_1(E)$, like those of other physical
quantities, bends up and saturates at $\vert E \vert = E_L$,
i.e. $ {{d \langle y_1 \rangle} \over dE} \rightarrow 0$.
Therefore, $P_{>}(E_m)$ decreases sharply as $E_m$ decreases to $E_L$.
Because $P_{<}(E_m)$ decreases linearly as $E_m$ increases to $E_L$,
we expect that $P(E_m)$ minimizes at
$E_m^{\rm min} \approx E_L \sim L^{-1/\nu}$, indicating the finite-size
cutoff, as discussed in greater detail in Appendix~\ref{app:modeling}.
Therefore, even though the CNN is trained at $E_m = E_m^{\rm min} \approx 0.125$
with seemingly perfect labels, it identifies as many conducting states for $E < E_m$
as for $E > E_m$ [see Fig.~\ref{fig2}(b)], which leads to the worst
performance.
The minimum of $P(E_m)$ indicates an indecision,
which is also supported by the equal number of conducting and localized states
below $E_m$, as shown in the lower-middle inset of Fig.~\ref{fig3}.

\begin{figure}
\centering
\includegraphics[width=8cm]{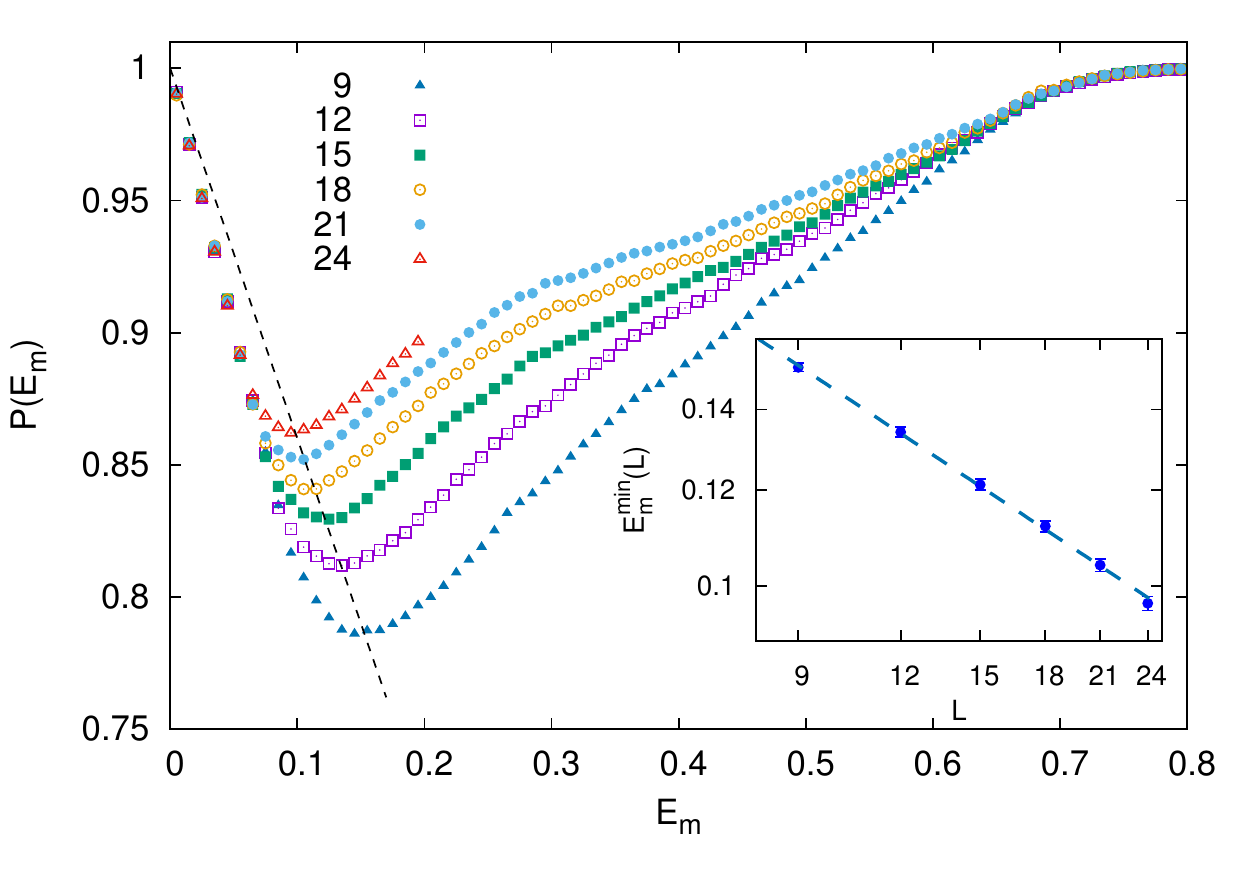} %
\caption{
(Color online.)
The performance curves for systems of size $L = 9$-24.
The minima of the curves can be extrapolated to $P(E_m) = 1$ at $E_m = 0$
in the thermodynamic limit.
The inset shows that the locations of the minima for various $L$ can be
fit by a power law $E_m^{\rm min} (L) \sim L^{-1/\nu}$ with
exponent $\nu = 2.22 \pm 0.04$.
}
\label{fig4}
\end{figure}

Figure~\ref{fig4} plots $P(E_m)$ as a function of $E_m$ for $L = 9$, 12,
15, 18, 21, and 24.
We apply polynomial fit to each curve and identified
the minimum $E_m^{\rm min}$ from the resulting fit.
Interestingly, $P(E_m^{\rm min})$ can be fit to a straight line
that extrapolates to 100\% accuracy at $E_m = 0$,
which is consistent with the fact that in the thermodynamic limit all
nonzero-energy states are localized.
We then plot $E_m^{\rm min}(L)$ as a function of $L$
in a double-logarithmic plot in the inset of Fig.~\ref{fig4}.
As expected, the data can be fit by a straight line, i.e. a power law
$E_m^{\rm min} (L) = a L^{-1/\nu}$, where $a = 0.410 \pm 0.008$
and $\nu = 2.22 \pm 0.04$.

\section{Conclusion and Discussion}
\label{sec:conclusion}

We train a deep CNN to study the quantum Hall plateau transition based on
the wave functions in a disordered lattice model.
With trial-and-error labelling for the conducting properties of the states,
the network can learn to classify the states with high but not perfect accuracy.
The imperfect classification reveals the critical fluctuations,
which allows us to quantify the range of the critical regime and
extract the localization length critical exponent.

The numerical determination of $\nu$ for plateau transition is known
to be difficult due to the existence of an irrelevant perturbation which
is close to marginal (or even marginal).~\cite{slevin09,obuse12,zhu18,puschmann18}
A very recent theoretical preprint even suggested that $\nu$ can keep increasing
as the renormalization group fixed point is approached.~\cite{zirnbauer18}
In this paper, we are able to extract $\nu = 2.22 \pm 0.04$
for the plateau transition for lattice sizes up to $L = 24$.
The result is robust against the change of network parameters
(see Appendix~\ref{app:robust} for detail).
The exponent is close to early numerical results $\nu = 2.4 \pm 0.1$
of the lattice model,~\cite{yang96,bhatt02}
which were {\it based on systems of similar sizes},
or to $\nu = 2.35 \pm 0.03$~\cite{huckestein95,evers08,slevin12}
in others models of the quantum Hall plateau transition.
Exponents that are larger than earlier results develop in lattice systems
only when their sizes are comparable to or greater than
$L = 100$,~\cite{zhu18,puschmann18}
which are beyond our current CNN training capacity.
Therefore, the reproduction of earlier results in this paper is
{\it not a defect of the method, but a vivid demonstration that the new method
is as robust as the conventional methods}.
However, future work is needed to boost the performance of the
trial-and-error scanning
to explore larger systems to reveal the larger exponent.
It would be interesting to explore whether the machine-learning method
can detect slowly decaying corrections to scaling.

\begin{figure}
\centering
\includegraphics[width=8cm]{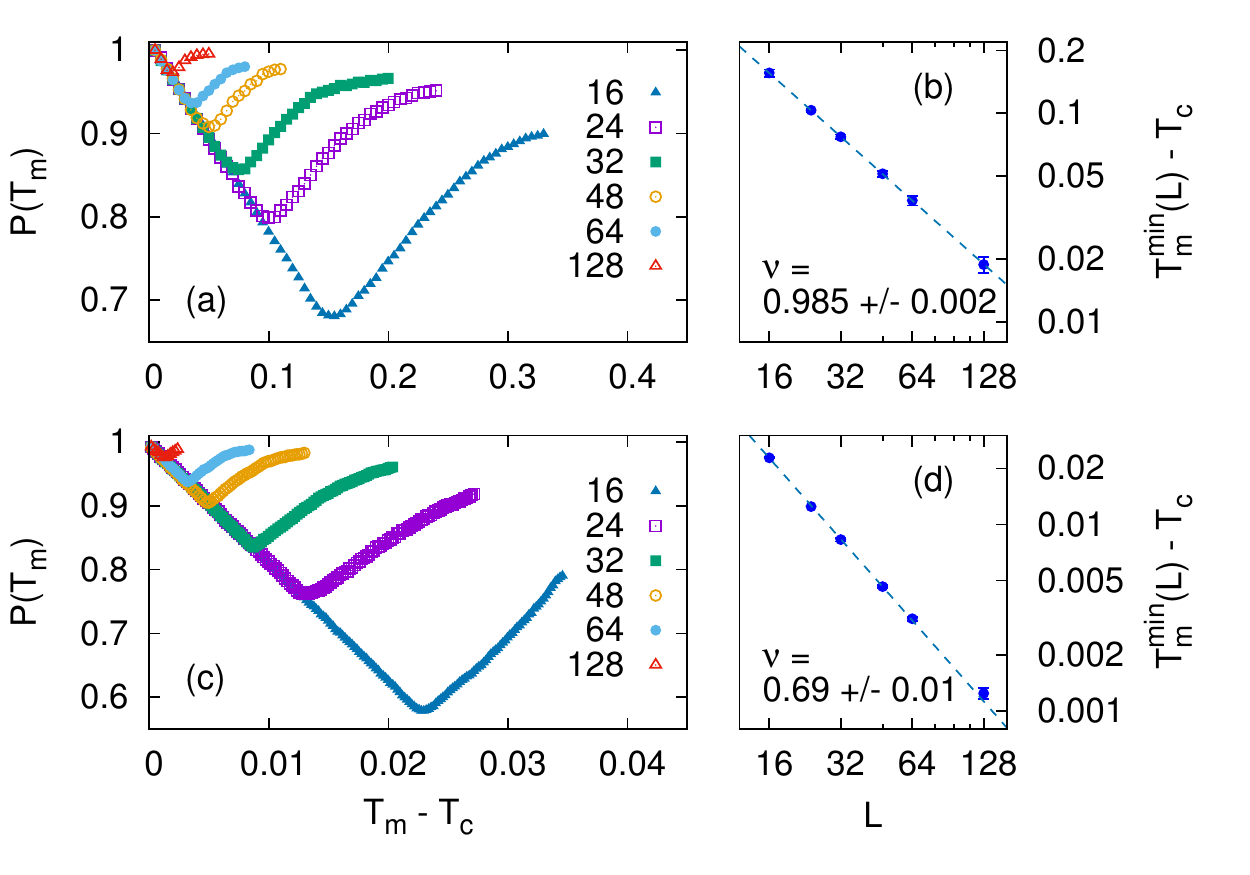} %
\caption{
(Color online.)
The performance curves for (a) the 2D Ising model and (c) the 4-state
Potts model on a square lattice.
The minima of the curves $T_m^{\rm min} (L) - T_c \sim L^{-1/\nu}$
can be used to extract critical exponent
(b) $\nu = 0.985 \pm0.002$ in the Ising model
and (d) $\nu=0.69\pm0.01$ in the Potts model.
}
\label{fig5}
\end{figure}

It is understandable that such a learning-from-data method is not
as efficient as conventional methods based on physical insights.
But the inefficiency is well compensated by its general applicability
to systems with drastically different physics.
For example, the machine learning based finite-size scaling method
can also be applied to thermal phase transitions.
Here, we briefly demonstrate its applicability to the 2D Ising and Potts models
in Fig.~\ref{fig5}.
When we feed the CNN with the spin configurations on the disordered side
of the thermal transitions,
we obtain similar V-shaped curves in the CNN performance and
the minima of the curves can be fitted to yield correlation length
critical exponent $\nu=0.985\pm0.002$ for the Ising model
and $\nu=0.69\pm0.01$ for the 4-state Potts model,
which are close to the exact results of 1 and 2/3,
respectively.
While several groups~\cite{Carrasquilla17,koch18,efthymiou18} have
estimated the critical exponents of the Ising model successfully,
the current paper proposes a versatile method that can be applied
to both electronic and spin models beyond the prototypical Ising model
{\it without changing the network architecture and training process}.
We leave the detailed discussion on the network parameters and
dataset preparation to Appendix~\ref{app:spin}.
We point out that in the two spin models we use the exact results of the critical temperatures.
Alternatively, one can either introduce $T_c$ as an additional fitting parameter in the finite-size scaling,
or determine $T_c$ by scanning data in a broader range with the confusion scheme.~\cite{Nieuwenburg17}

\begin{acknowledgments}
This work was supported by the National Natural Science Foundation
of China through Grant No. 11674282 and the National Basic Research Program of China
through Project No. 2015CB921101.
XW also thanks the hospitality of
the Yukawa Institute for Theoretical Physics
during the workshop ``Novel Quantum States in Condensed Matter 2017''
during which the manuscript was being finalized and presented.

\end{acknowledgments}

\appendix

\section{The convolutional neural network}
\label{app:network}

In this appendix, we illustrate the architecture and training of the convolutional neural network
in the context of the quantum Hall plateau transition.
The network structure and the learning scheme can be straightforwardly
generalized to other models, such as the 2D Ising model and the Potts model
to be discussed in Appendix~\ref{app:spin}.

\subsection{Neural network architecture}
\label{app:architecture}

\begin{figure}
\begin{center}
\includegraphics[width=7.5cm]{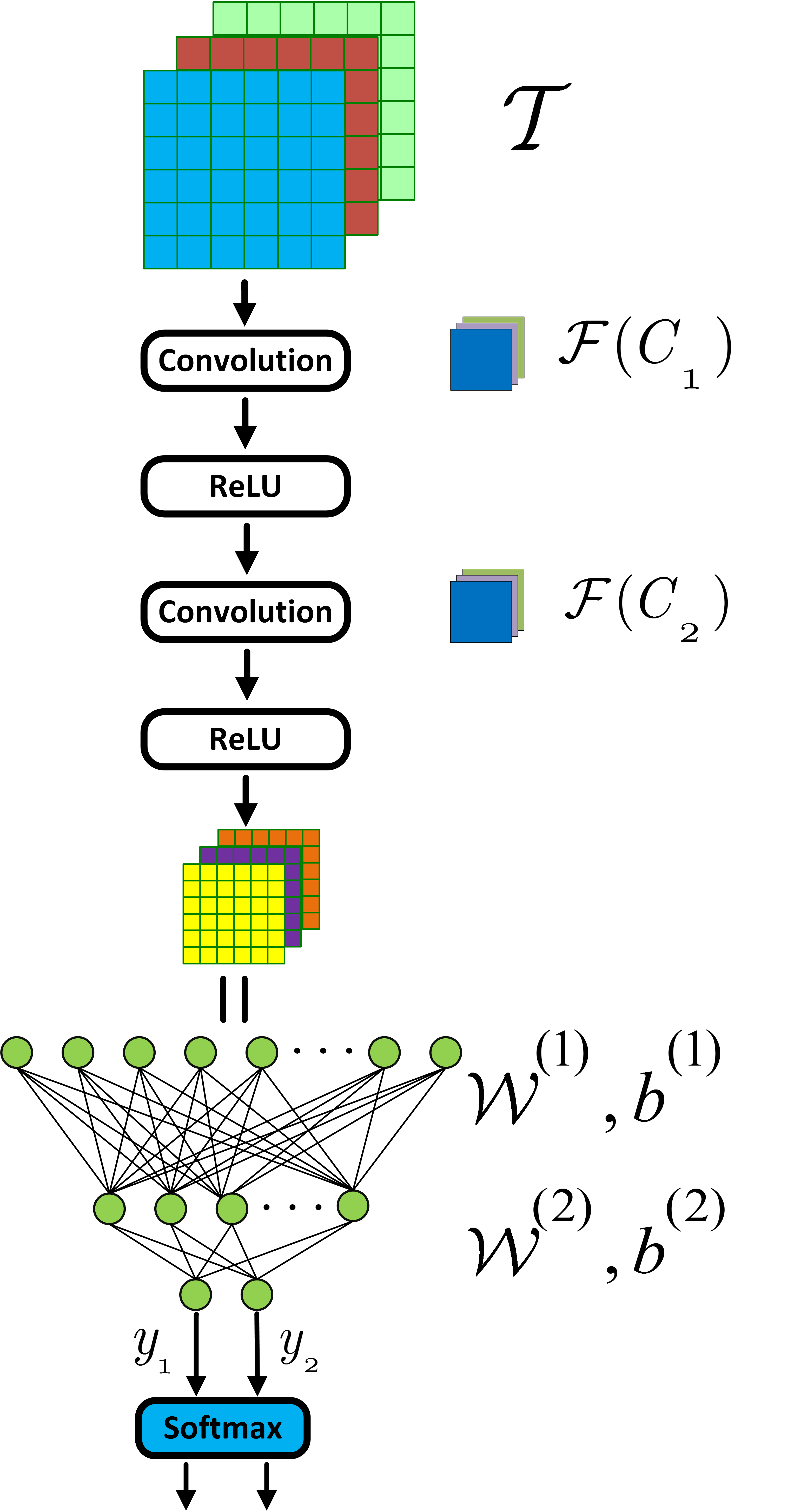}
\end{center}
\caption{(Color online.)
Architecture of the CNN model used in the present study. }
\label{FIG-PEND}
\end{figure}

The CNN model we use is conceptually similar to the LeNet construction that is used
for recognizing characters.~\cite{lecun98}
It consists of two convolutional layers with ReLU activation and two fully connected layers,
as illustrated in Fig.~\ref{FIG-PEND}.
For normalized eigenstates $\vert \psi \rangle = \sum_i c_i \vert i \rangle$,
the architecture of the model $\mathcal{N}_{\rm CNN}$ is as follows.\\
\begin{equation}
\begin{scriptsize}
\left[
  \begin{array}{l}
\mathcal{T}=\left \lbrace \left \lbrace \left \vert c_{i} \right \vert^4 \right \rbrace:
{\rm Wave ~functions ~on ~an ~} L \times L ~ {\rm lattice} \right \rbrace  \\
\newline \\
\downarrow\left\{ \begin{array}{l}
\rm 1st~Convolution_{[N_{\it f}^2 - filter,(s,s)-stride,C_1-channels]} \\
\newline \\
\rm ReLU~activation\\
\newline \\
\rm 2nd~Convolution_{[N_{\it f}^2-filter,(s,s)-stride,C_2-channels]}\\
\newline \\
\rm ReLU~activation\\
\newline \\
\rm Flatten \\
\end{array}
\right.\\
\newline \\
\mathbb{R}^{L^2/s^4\times C_2}\\
\newline \\
\downarrow\left\{ \begin{array}{l}
\rm 1st~Fully~connected\\
\newline \\
\rm 2nd~Fully~connected\\
\newline \\
\rm Softmax\\
\end{array}
\right.\\
\newline \\
\mathbb{R}^{N}=\mathcal{O}
\end{array}
\right]
\end{scriptsize}
\label{CNN_architecture}
\end{equation}
$\mathcal{T}$ is the training data set and $\mathcal{O}$ is the output set.
In the model, we identify only two classes, conducting states and localized states, so $N=2$ for the output.
We train the CNN model with the following parameters,
\begin{equation}
\rm N_{\it f}=3,~s=1,~C_1=16,~C_2=18.
\end{equation}
The first convolutional layer has 16 feature maps.
With stride 1 and zero padding, the size of each feature map is $L \times L$.
Each unit in the feature maps is filtered from a $3 \times 3$ neighborhood in the input.
Altogether, the layer contains $16 \times (3 \times 3 + 1) = 160$ trainable parameters
(including 16 bias parameters).
The convolutional layer is followed by a layer of non-linear activation units ReLU with a non-saturating activation function
\begin{equation}
f(x) = \left \{ \begin{array}{ll}
x, &x > 0 \\
0, &x \leq 0
\end{array}
\right . .
\end{equation}
The second convolutional layer has 18 feature maps.
With stride 1 and zero padding, the size of each feature map is, again, $L \times L$.
Each unit in the feature maps is also filtered from a $3 \times 3$ neighborhood in the input.
The layer contains $18 \times [16 \times (3 \times 3) + 1] = 2,610$ trainable parameters.
After a second ReLU layer, the 18 feature maps are flattened into a one-dimensional array of
$18 L^2$ nodes.
The nodes are then fully connected to a layer of 256 nodes with ReLU activation.
The fully connected layer has $256 \times (18 L^2 + 1)$ trainable parameters
(including 256 bias parameters).
The 256 nodes are fully connected to 2 output nodes with $2 \times (256 + 1) = 514$
trainable parameters.
If we denote the values of the 2 nodes as $y_1$ and $y_2$,
the corresponding softmax (or logistic) outputs
\begin{equation}
{\rm softmax}(y_i) = \frac{e^{y_i}}{e^{y_1} + e^{y_2}}
\end{equation}
are the probabilities that the CNN associates with a state
being localized or conducting.

We have two sets of parameters $\mathcal{F}$ and $\mathcal{W}$ (we neglect the corresponding bias parameters
in illustration for simplicity):
\begin{itemize}
\item filters $\mathcal{F} \equiv \left \{ \mathcal{F}_{ij}^a(C_1),\mathcal{F}_{kl}^{bc}(C_2) \right \}$ for convolutional layers,
where $a,b=1,\cdots,C_1$, $c=1,\cdots,C_2$, and $i, j, k, l=1,\cdots,N_f$, and
\item weights $\mathcal{W} \equiv \left \{
\mathcal{W}^{(1)}_{mn}, \mathcal{W}^{(2)}_{pq} \right \}$ for fully connected layers, where
$m=1,\cdots,L^2 \times C_2$, $n, p =1,\ldots,256$, and $q=1,2$.
\end{itemize}

To avoid overfitting, we implement dropout for the fully-connected layers and the dropout ratio
is chosen to be 0.5.

\subsection{Preparation of the training data set}
In order to prepare the training data set, we first prepare the set of data-label pairs
\begin{displaymath}
\left \{
\left ( \left \lbrace \left \vert c_{I,i} \right \vert^4 \right \rbrace, L_I \right )
\left \vert
\begin{array}{lll}
L_I = & (1,0), & {\rm if} ~\vert E_I \vert < E_m \\
& (0, 1), & {\rm otherwise}
\end{array}
\right .
\right \}.
\end{displaymath}
Here, $E_I$ is the energy of $I$th wave function $\vert \psi_I \rangle = \sum_i c_{I,i} \vert i \rangle$,
while $L_I$ is the one-hot representation of the corresponding label imposed by $E_m$.
We prepare $N_{\rm state}$ states by diagonalizing the Hamiltonian with different realizations of
random potential.
$E_m$ is the moving postulation of the energy that separates localized states from conducting ones.
We separate the data set into a training set and a test set. In order to avoid the vanishing gradients problems,
we set $\langle\left \vert c_{i} \right \vert^4 \rangle=0$ to make sure our data is zero-centered.

As we discuss in Sec.~\ref{sec:model}, we want to estimate the width of the density of conducting states.
However, we do not know which states are conducting {\it a priori}. So we introduce a trial-and-error
approach. We assume that there is a characteristic energy scale $E_m$, below which the states are
conducting and above which the states are localized.
We want to use the CNN to select, among all possible $E_m$, one energy scale that can
characterize the critical fluctuations.
To proceed, we label all the states with energy $|E| < E_m$ as conducting states [$L_n = (1,0)$]
and the rest as localized states [$L_n = (0,1)$] for a given $E_m$.
We train the CNN model $\mathcal{N}_{\rm CNN}(E_m)$ with the training data and
evaluate the accuracy rate of $\mathcal{N}_{\rm CNN}(E_m)$ with the separate test data.
The resulting rate is referred to as the performance $P(E_m)$ of the CNN.
We scan $E_m$ within the entire energy range and Sec.~\ref{sec:results} explains how we obtain
from $P(E_m)$ the characteristic energy scale for various system sizes.

\subsection{Training}
For a given $E_m$, our training process can be represented by
\begin{equation}
F \left (\mathcal{N}_{\rm CNN}(E_m) \right ): \mathcal{T}\rightarrow \mathcal{O}.
\label{trainig_qhe}
\end{equation}
We define the cross entropy for a batch of $N_b$ states
\begin{equation}
\mathcal{E}
=- \sum_I^{N_b} \sum_{i=1}^{2}L_{I,i} \log L_{I,i}^{\rm CNN},
\label{loss_qhe}
\end{equation}
where $\left (L_{I,1}^{\rm CNN}, L_{I,2}^{\rm CNN} \right )$ is the softmax output for the $I$th state,
whose desired label imposed by $E_m$ is $\left (L_{I,1}, L_{I,2} \right )$ in the one-hot representation.
The loss function we train to minimize is the sum of the cross entropy and
the L2-regularization of the weights $\mathcal{W}$ associated with the fully connected layers, i.e.
\begin{equation}
\mathcal{E} + \beta \left [
\sum_{m}^{L^2 \times C_2} \sum_n^{256} \left \vert \mathcal{W}^{(1)}_{mn} \right \vert^2 +
\sum_{p}^{256} \sum_q^2 \left \vert \mathcal{W}^{(2)}_{pq} \right \vert^2
\right ],
\label{l2_qhe}
\end{equation}
where the regularization weight $\beta$ is chosen to be $5\times 10^{-4}$.

For each system size, we choose 300,000 states for training and 35,000 for testing.
We vary $E_m$ in the energy range $(0, 0.8)$ with interval $\delta E = 0.01$ and
use the Adam optimizer (tf.AdamOptimizer in TensorFlow v1.0)
to update $\mathcal{F}$ and $\mathcal{W}$.
We choose batch size $N_b = 2000$ and learning rate $\eta = 10^{-3}$.
The weights $\mathcal{W}$ are initiated from a truncated normal distribution with zero mean and
standard deviation $0.06$.
The initial bias parameters are chosen to be 0.1 for all layers.
We perform at least 800,000 iterations for each $E_m$ to obtain the
V-shaped performance curve we discuss in Sec.~\ref{sec:results}.

\subsection{Weight initialization}

We find that the efficiency of the training process depends crucially on the initialization
of the weights $\mathcal{W}$.
If we randomly initialize the weights, it takes about 800,000 iterations
for the test accuracy to saturate.
Our observations suggest that during the long training process
the amplitudes of most of the $\mathcal{W}$ reduce to be
significantly smaller than the initial standard deviation 0.06.
However, if we initialize the $\mathcal{W}$ with a much narrower truncated normal distribution,
the training is unstable and the resulting test accuracy has large fluctuations as $E_m$ varies.

Alternatively, we can scan $E_m$ from 0 to 0.8 and
store the optimal $\mathcal{F}$ and $\mathcal{W}$ for each $\mathcal{N}_{\rm CNN}(E_m)$
for the initialization of the adjacent $\mathcal{N}_{\rm CNN}(E_m + \delta E)$.
We find that a mere 80,000 iterations are enough to reach the saturation of the test accuracy.
We also find that both the random initialization and the optimal (at the adjacent $E_m$) initialization
generate almost the same performance curve as shown in Fig. 4.
The latter is computationally more efficient.

\subsection{Random errors in $E_m^{min}$}

Our performance curves represent the maximum test accuracy of the CNN
for the varying $E_m$.
We apply polynomial fit for each system size to obtain the location of
the minimum $E_m^{\rm min}$.
To estimate the random errors of $E_m^{\rm min}$,
we trained the CNN up to 500 times with independent random initialization of
filters, weights and biases of the network
so that we obtain a distribution of the minimum location,
from which we can estimate the random errors of $E_m^{\rm min}$.
We further obtained the critical exponent $\nu$ and its error bar
by fitting $E_m^{\rm min}$ with error bars to the power-law scaling form
$E_m^{\rm min}  = a L^{-1/\nu}$.

\section{Modeling the performance curve}
\label{app:modeling}

The V shape of the training curve is not easy to understand at first glance.
As we mentioned in Sec.~\ref{sec:results}, one naively expects that
the data should be consistent with the labels when $\xi(E_m) \sim L$,
which, apparently, would lead to a maximum.
But this is incorrect.
One needs to consider the fluctuations in physical quantities,
as well as in the features identified by the CNN.
Here we provide more details on the cause of the V-shaped performance curve,
and the meaning of its minimum at $E_m^{min}$.

\begin{figure}
\begin{center}
\includegraphics[width=7.5cm]{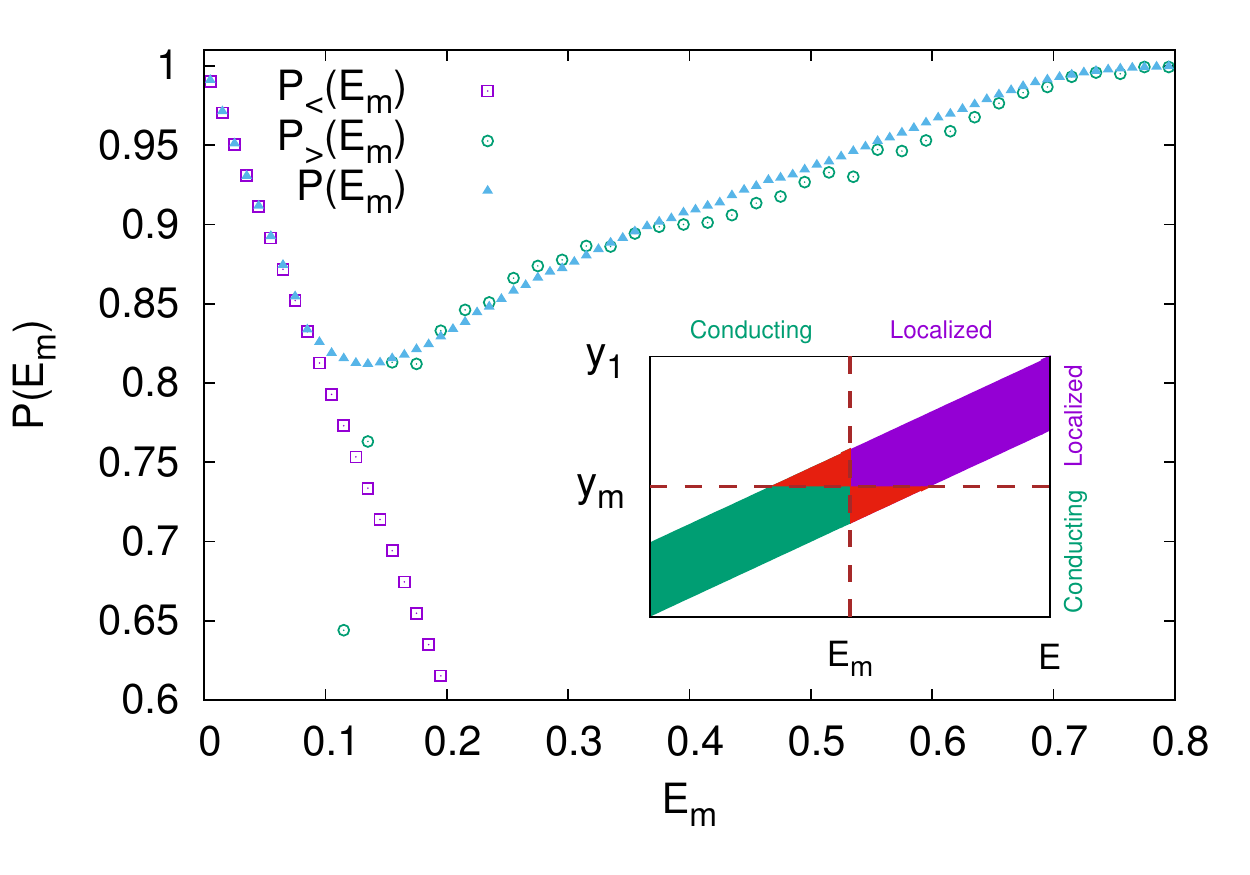}
\end{center}
\caption{(Color online.)
The performance curve $P(E_m)$ and its low-energy approximation
$P_{<}(E_m)$ and high-energy approximation $P_{>}(E_m)$ for systems of size
L = 12. The minima of $P(E_m)$ can be estimated by letting $P_{<}(E_m) = P_{>}(E_m)$.
The shape factor $\zeta = 0.7$ is used for calculating $P_{>}(E_m)$.
In the inset we illustrate the fluctuations of the CNN output
as a function of energy.
The input label of a state is determined by its energy
while the classification of the state is by the CNN output $y_1$ (and $y_2$). For states with fluctuating features, the classification
cannot agree with the input labels in the red regions, leading to an
imperfect performance.
}
\label{fig2app}
\end{figure}

As $E_m$ increases from zero, $P(E_m)$ first decreases
because of the increasing inconsistent labels.
For small $E_m$ where $L < \xi(E_m)$,
we assume in the labelling scheme that states with energy $|E| < E_m$ are conducting,
but the smallness of their number misleads the CNN to conclude that all states
are localizing,
which has a better performance.
So $1 - P_{<}(E_m)$ is the percentage of the states with $|E| < E_m$,
\begin{equation}
Q_{<} \equiv 1 - P_{<}(E_m) = \int_{-E_m}^{E_m} \rho(E) dE \approx 2 E_m \rho(0),
\end{equation}
where $\rho(E)$ is the normalized density of states.
Note this result is indenpendent of system size $L$.
We plot $P_{<}(E_m)$ in Fig.~\ref{fig2app} and find that $P_{<}(E_m)$ agrees
perfectly with the training result $P(E_m)$ for $E_m \leq 0.9$ (with $L = 12$).
In this range, $\rho(E_m) \approx \rho(0)$ as the density of states is flat,
so we can also write
\begin{equation}
\label{eq:q1}
P_{<}(E_m) = 1 -  2 E_m \rho(E_m),
\end{equation}

On the other hand, for sufficiently large $E_m$,
nonzero $1 - P(E_m)$ comes from the fluctuations of the feature recognized by the CNN.
We assume a feature $y$ recognized by the CNN, e.g., $y_1$ of the output neuron,
can be represented by a distribution with mean $y(E; E_m)$ with
fluctuations or variance $\delta y(E; E_m)$.
Figure~\ref{fig2app}
plots $y(E; E_m) - y(0; E_m)$ and $\delta y(E; E_m)$ for $E_m = 0.105$,
0.225, and 0.345.
We emphasize that the dependence of the distribution on $E_m$
is not significant so we can drop $E_m$ and simply write
$y(E)$ and $\delta y(E)$, at least for $E_m$ to the right of the minimum of
the V-shaped $P(E_m)$.

To estimate $1 - P(E_m)$, we can approximate $y$ versus $E$ by
a smooth varying curve with constant broadening $\delta y$,
on which the data is uniformly scattered, as illustrated in the inset
of Fig.~\ref{fig2app}.
The input label of a state is determined by its energy
while the classification of the state is by the CNN output $y_1$
and $y_2$. In the two-class situation, it is normally sufficient to
explore $y_1$ only in the modeling, as we find $y_2 \approx - y_1$ in practice.
Now, the states whose conducting nature are recognized incorrectly
by the CNN are distributed in the two red triangles in the inset of Fig.~\ref{fig2app}.
Their percentage is
\begin{equation}
Q_{>} \equiv 1 - P_{>}(E_m) \approx 2 \times 2 \times {1 \over 4} \rho(E_m) \delta E,
\end{equation}
where the width of the triangle $\delta E \approx \delta y / (dy/dE)$
can be evaluated at $E_m$.
The first 2 comes from two fictitious mobility edges $\pm E_m$,
and the second 2 from the sum of two triangles, and
the 1/4 means that only a quarter of the states in the energy range
$\delta E$ are recognized incorrectly.
We can introduce a shape factor $\zeta$
to account for the real distribution of $y$ such that
\begin{equation}
\label{eq:q2}
P_{>}(E_m) = 1 - \zeta \rho(E_m) \left . {{\delta y} \over dy/dE} \right \vert_{E=E_m}
\end{equation}
We plot $P_{>}(E_m)$ in Fig.~\ref{fig2} with $\zeta = 0.7$,
which is computed from $y(E; E_m)$, $\delta y(E; E_m)$,
and $\rho(E_m)$ according to Eq.~(\ref{eq:q2}), and find that
$P_{>}(E_m)$ agrees well with the training result $P(E_m)$ for $E_m \geq 0.16$.
In this range, $P_{>}$ increases with $E_m$ due to the decreasing
$\rho(E_m)$, while both $\delta y$ and $dy/dE$ saturate.

\begin{figure}
\begin{center}
\includegraphics[width=7.5cm]{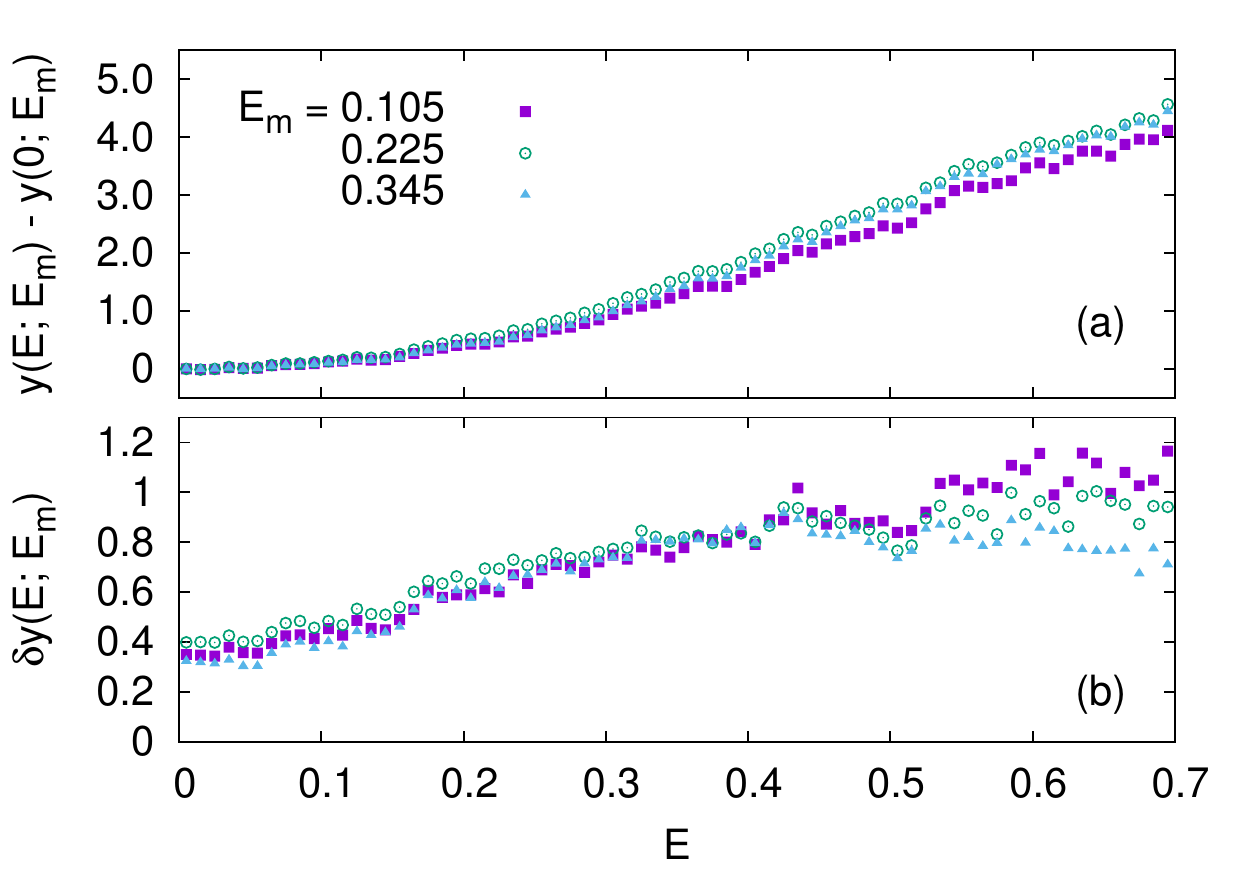}
\end{center}
\caption{(Color online.)
(a) Shifted mean $y(E; E_m)$ and (b) its variance $\delta y(E; E_m)$ of the CNN output
$y_1$ for $E_m = 0.105$, 0.225, and 0.345. The lattice size is $L = 12$. }
\label{fig3app}
\end{figure}

The V-shaped $P(E_m)$ is therefore the smooth interpolation of
the two limiting behavior, i.e., $P_{<}(E_m)$ for small $E_m$ and
$P_{>}(E_m)$ for large $E_m$.
Let us focus on the crossover from small $E_m$ to
large $E_m$.
It is not difficult to observe in Fig.~\ref{fig3app} that $y$ saturates
below the crossover.
As for other physics quantities, the saturation is a finite-size effect;
when $L < \xi(E_m)$, the CNN can no longer detect
the increasing localization length.
In this regime $dy/dE$ approaches 0,
so $P_{>}(E_m)$ drops below $P_{<}(E_m)$,
which is linear in $E_m$.
Once $L > \xi(E_m)$, $dy/dE$ increases sharply from zero,
so $P_{>}(E_m)$ becomes greater than $P_{<}(E_m)$.
The crossover $E_m^{min}$ is, thus, the energy scale
when $L \sim \xi(E_m)$.
In other words, it detects the size cutoff in a finite system.

It is interesting to note that there is a bulge in $P_{>}(E_m)$
around $E_m = 0.3$.
The origin of it can be traced to the competition of
the increasing fluctuations $\delta y$ in the CNN output and
the decreasing density of states as $E_m$ increases.
The reminiscent of the bulge can be found in the
V-shaped performance curves in Fig.~\ref{fig4},
but it is irrelevant to criticality and
not expected to develop into a W shape
by further training.

\section{Robustness of results}
\label{app:robust}

We use CNN to extract the critical exponent $\nu$ for the integer quantum plateau transition in the Hofstadter lattice model.
In order to test the robustness of our results, we vary the network architecture, the activation function, the size and the number of filters.
We show that the successful application of the method to the plateau transition requires at least two CNN layers, as shown in the CNN architecture in Fig. 1.
The form of the activation function is not crucial as long as we avoid the Sigmoid function, which is known to cause the vanishing gradient problem in the backpropagation algorithm.
Our results are stable when the filter size and the number of filters are changed.

\begin{figure}
\begin{center}
\includegraphics[width=7.5cm]{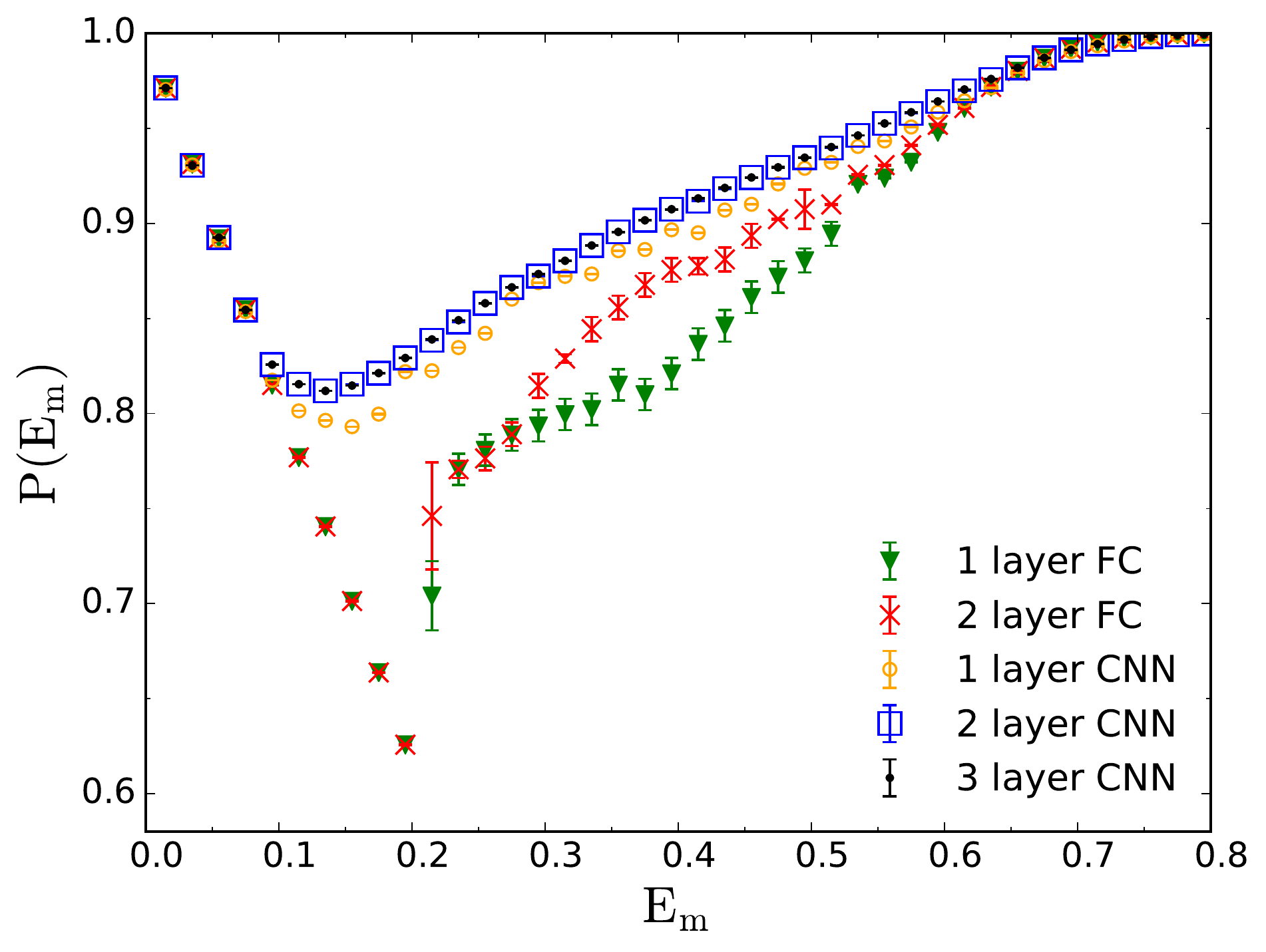}
\end{center}
\caption{(Color online.)
The performance curve obtained by different CNN network layers for the Hofstadter model.}
\label{FIG-qhe_diff_layer}
\end{figure}

\subsection{The effects of different layers}

We consider five different neural network architectures.
With increasing complexity, they are
(1) one layer of fully connected (FC) network,
(2) two layers of FC network,
(3) one layer of CNN followed by two layers of FC network,
(4) two layers of CNN followed by two layers of FC network, and
(5) three layers of CNN followed by two layers of FC network.
For comparison, we fix the system size to be $L = 12$. Figure~\ref{FIG-qhe_diff_layer} compares the performance of the five cases.

In the absence of CNN layer(s) [cases (1) and (2)], the performance curves show a sudden jump near $E_m = 0.22$.
According to the inset of Fig. 2, this is when the states at $E = 0$ are mostly identified as conducting.
The straight line for $E_m < 0.22$ implies that almost no conducting states are identified. The jump mimics a first-order transition. Therefore, FC layers only fail to identify the critical fluctuations.

With one layer of CNN [case (2)], the neural network improves its performance significantly. The performance curve is reasonably smooth.
This means that one layer of CNN is able to identify the critical fluctuations, though with poor performance, such that $E_m^{min}$ has a large error bar.

For two and three layers of CNN [case (4) and (5)], we find almost identical  V-shaped performance curves. We choose $C_1=16$, $C_2=16$, and $C_3=8$ for the three layers in case (5), which is more time-consuming than case (4).
Therefore, in practice, we choose the two-layer CNN to study the critical exponent in Sec.~\ref{sec:results}.

\subsection{The effects of different non-linear active function}

In the main text, the non-linear active function is chosen to be ReLU,
which is commonly used in effective training of deep neural networks.
The choice is not unique.
For comparison, we also use exponential linear unit (ELU), Leaky ReLU, and Sigmoid activation functions in our CNN model.

\begin{figure}
\begin{center}
\includegraphics[width=7.5cm]{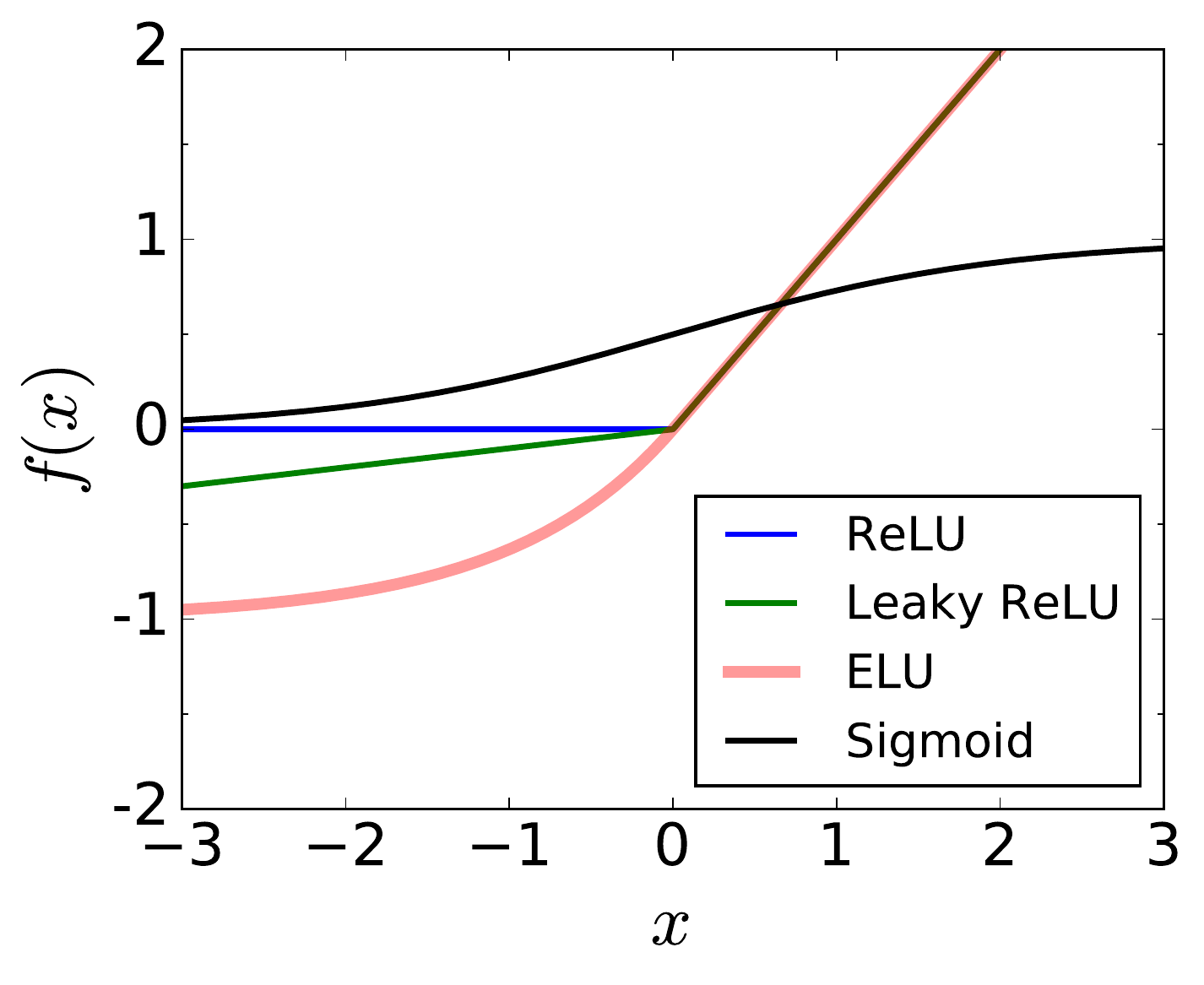}
\end{center}
\caption{(Color online.)
Comparison of four commonly used activation functions: ReLU, Leaky ReLU, ELU, and Sigmoid functions.}
\label{FIG-diff_actv_fucn}
\end{figure}

These non-linear activation functions are defined as follows.
\begin{enumerate}
\item Leaky ReLU, which introduces a small positive gradient when the unit is not active to avoid the vanishing gradient problem,
\begin{equation}
f(x) = \left \{ \begin{array}{ll}
x, &x > 0 \\
\alpha x, &x \leq 0
\end{array}
\right . ,
\end{equation}
where $\alpha<1$ is a small non-negative constant. We choose $\alpha=0.01$.

\item ELU, whose mean activations is closer to zero,
\begin{equation}
f(x) = \left \{ \begin{array}{ll}
x, &x > 0 \\
a (e^{x}-1), &x \leq 0
\end{array}
\right . ,
\end{equation}
where $a \geq 0$ is a non-negative constant to be tuned.

\item Sigmoid, which is the classic non-linear activation function,
\begin{equation}
f(x) = \frac{1}{1+e^{-x}}.
\end{equation}
Sigmoid function saturated in both directions and suffer more from the vanishing gradient problem.

\end{enumerate}

\begin{figure}
\begin{center}
\includegraphics[width=7.5cm]{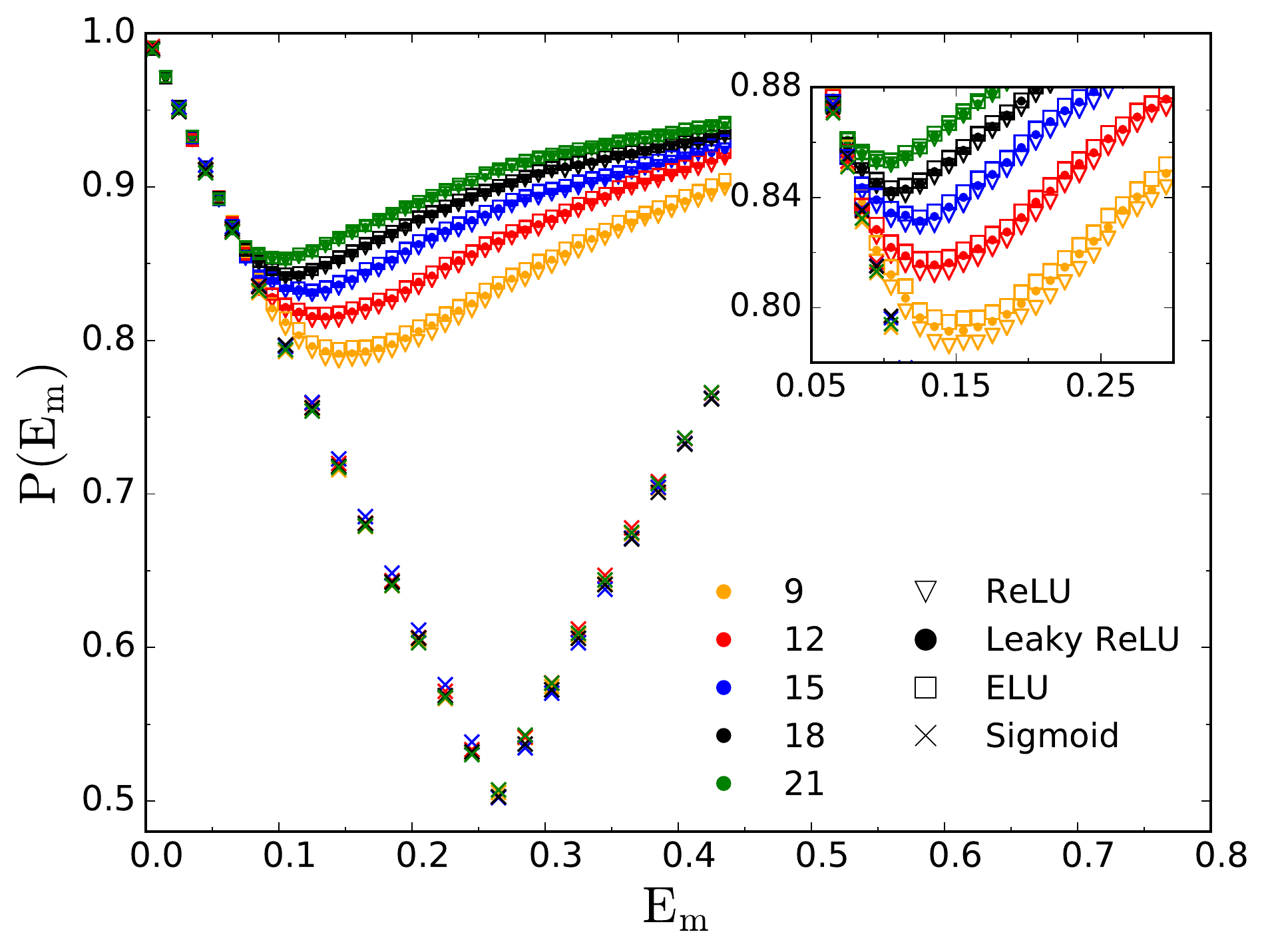}
\end{center}
\caption{(Color online.)
The performance curves obtained with different non-linear activation functions for the Hofstadter lattice model. The inset amplifies the curves near their minima.}
\label{FIG-qhe_diff_actv_fucn}
\end{figure}

The three activation functions are compared, together with ReLU, in Fig.~\ref{FIG-diff_actv_fucn}.
In our comparison study we replace all activation functions but keep all other parameters unchanged as in the ReLU case.
Fig.~\ref{FIG-qhe_diff_actv_fucn} shows the different V-shaped performance obtained by different non-linear active functions for system size $L$ from 9 to 21. With ReLU, Leaky ReLU, and ELU, we obtain similar results.
We find that, in terms of performance, ELU and Leaky ReLU are slightly better than ReLU, but Sigmoid activation is significantly worse.
Therefore, one needs to use activation functions that do not suffer significantly from the vanishing gradient problem.
The critical exponent obtained from ReLU, Leaky ReLU, and ELU activations agrees within error bars.

\subsection{The effects of the size and number of filters}

In Sec.~\ref{sec:results} we use 3$\times$3 filters for our CNN.
We also train the CNN with larger 5$\times$5 and 6$\times$6 filters.
We keep other parameters the same as in Sec.~\ref{sec:model}.
The trainings yield $\nu=2.23\pm0.08$ for the 5$\times$5 filters and
$\nu=2.25\pm0.08$ for the 6$\times$6 filters.
Therefore, the smaller 3$\times$3 filters perform as good as the larger ones.

\begin{figure}
\begin{center}
\includegraphics[width=7.5cm]{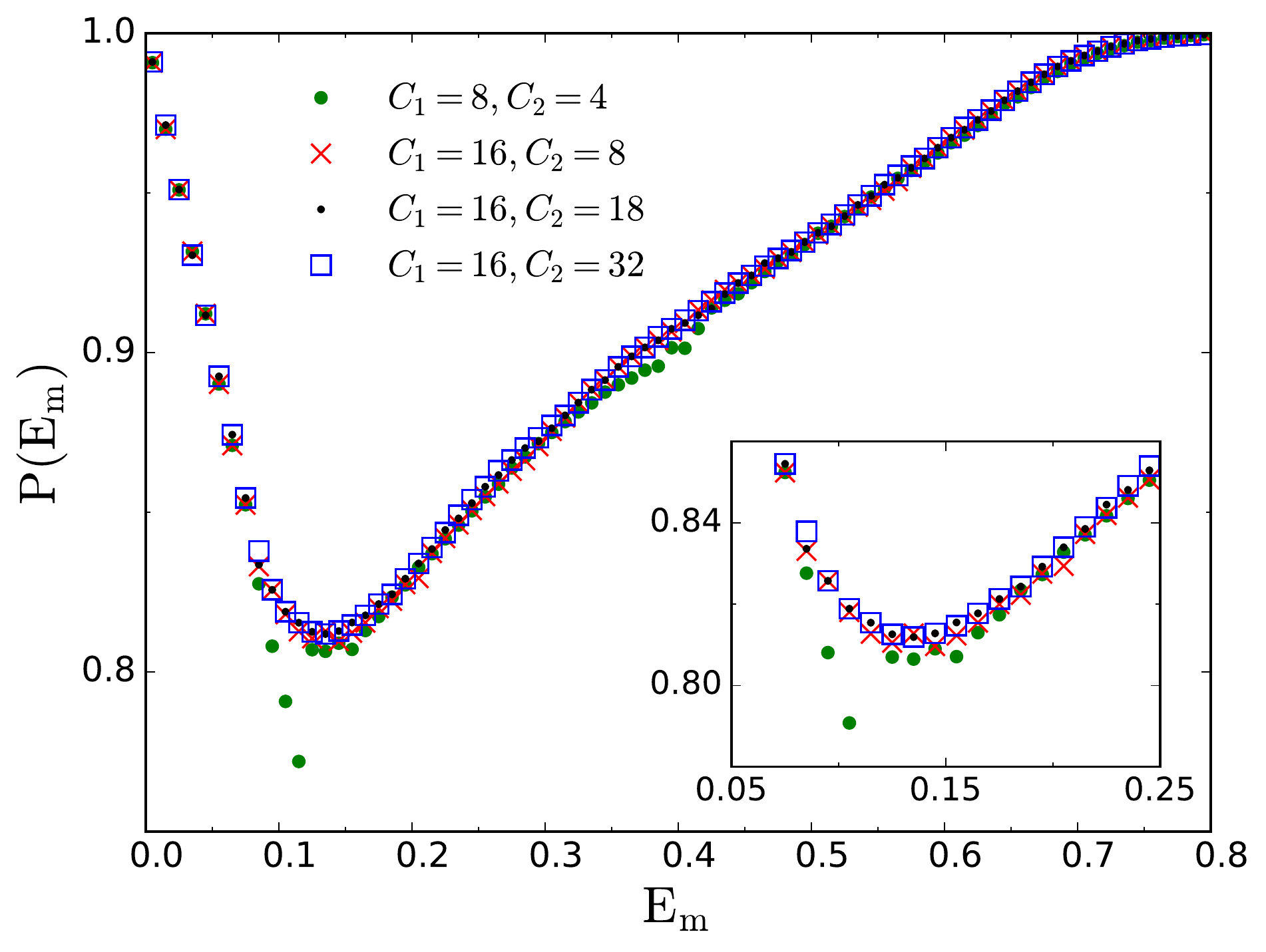}
\end{center}
\caption{(Color online.)
The performance curves obtained with different number of filters for the Hofstadter model. The inset amplifies the curves near their minima.}
\label{FIG-qhe_diff_channel}
\end{figure}

On the other hand, the filter number in each layer needs to be sufficiently large. Figure~\ref{FIG-qhe_diff_channel} compares the performance curve for the two-layer CNN with different combinations of filter numbers in the two layers.
We use 3$\times$3 filters and the system size is $L = 12$.
For $C_1=8$ and $C_2=4$, an abrupt jump develops at the minimum, indicating that the fluctuations are not well captured.
The performance is generally better for more filters, but the improvement is not significant beyond $C_1=16$ and $C_2=8$.
We choose $C_1=16$ and $C_2=18$ in Sec.~\ref{sec:results} to balance the performance and the time consumption.

\section{Application to other models}
\label{app:spin}

In this section, we discuss the generalization of the method using CNN to extract critical exponents.
We have trained the CNN to learn the Ising model and the four-state
Potts model on a square lattice.
We have presented the V-shaped performance curves and the corresponding finite-size scaling of the minima in Fig. 5.
Here we explain the details in data preparation and training.

Before we start, let us point out the major differences between the plateau transition and the Ising/Potts transitions from machine learning point of view.
\begin{enumerate}
\item The plateau transition is a quantum phase transition.
As the Fermi energy sweeps across the band center, the Hall conductance increases from 0 to 1.
The natural input to the CNN for the plateau transition is a quantum wave function or local electron density.
Each wave function can be labelled by its corresponding eigenenergy.
The Ising and Potts transitions are thermal transitions, whose critical temperatures are rigorously known.
The obvious input for the spin models are local spin configurations.
However, when we generate spin configurations by Monte Carlo (MC) simulations,
there are multiple configurations at the same temperature.
The configuration to configuration fluctuations are the dominating physics
the CNN needs to understand.
\item It remains difficult to find an obvious order parameter for the localization problem, while there are well-defined local order parameters in the Ising and Potts models.
However, having simple order parameter may be a disadvantage for the critical exponent extraction, because it is easy for the machine to achieve perfect performance. As a matter of fact, we made breakthrough first in understanding the plateau transition, rather than the simpler Ising model.
\end{enumerate}
While the main architecture and the training process remain unchanged, we introduce minor changes to the network architecture and data preparation
to take these differences into account.

Take the two-dimensional Ising model on a square lattice of size $L \times L$ with ferromagnetic couplings as an example.
We use Metropolis sampling in the MC simulation to prepare data of Ising spin configurations from a cold start with all spin configuration equal to $+1$ for temperature $T > T_c$.
Following the practice in the plateau transition, we define a crossover temperature $T_m$, such that we label the data with $T_c < T < T_m$ as ferromagnetic and data with $T > T_m$ as paramagnetic.
Notice that we are not using the data below $T_c$ in the ferromagnetic phase, because (1) it is sufficient to obtain the correlation length critical exponent above $T_c$ and (2) the fluctuations in the ferromagnetic phase is weak compared to the dominant order parameter information for the CNN to learn quantitatively.
The key to capture the thermal fluctuations here is to note that the CNN needs to read the deviations of the configutrations from their average.
In contrast, the quantum fluctuations in the lattice model is embedded in wave functions, each of which has a unique energy (neglecting accidental degeneracy).
Therefore, we group a fixed number $N_c$ of spin configurations at the same temperature into a single (collective) sample of input, with different configurations as multiple channels.

In practice, we sample the spin configurations from $T_{min}=2.27$ to $T_{max}=2.8$ with temperature interval $\delta T=0.002$.
For each temperature we collect 20,000 configurations.
To avoid the vanishing gradient problem we subtract the magnetization per spin such that the mean spin of each configuration is zero.
We draw randomly from the 20,000 configurations at each temperature to prepare $N_s = 800$ samples of CNN input, each contains $N_c = 180$
channels of configurations.
The choice of $N_c$ optimizes the performance on our computer hardware
and is believed not to affect the final result.
Therefore, we have a total of $N_{s}(T_{max}-T_{min})/\delta T=212$,000 samples of CNN input for each system size, among which we select
170,000 samples for training and another 10,000 for testing.

The learning scheme is the same as for the plateau transition.
We vary $T_m$ in the temperature range between $T = 2.27$ and 2.8 with interval
$\Delta T=0.004$.
For a given $T_m$, the training process follows the same prescription: Eqs.~(\ref{trainig_qhe}), (\ref{loss_qhe}), and (\ref{l2_qhe}).
In the Ising case, the large size of the multi-channel data in each input sample is compensated by the ease of training.
We find it sufficient to use one convolutional layer and two fully connected layers.
We choose $C = 4$ channels of $L \times L$ CNN filters and apply ReLU activation function after convolution.
Through the two fully connected layers we first reduce the number of nodes to 64, then to 2 softmax outputs.
We use batch size $N_b=100$ and learning rate $\eta=10^{-3}$.
For each $T_m$, we test the performance after 30,000 iterations.
The resulting V-shaped performance curves for system size $L=16$, 24, 32, 48, 64, and 128 are plotted in Fig.~5(a).
From the minima of the curves, we perform finite-size scaling as in Fig.~5(b) and obtain the correlation length critical exponent
$\nu = 0.985 \pm 0.002$.
The result compares favorably to the exact value $\nu = 1$.

We also study the 4-state Potts model on a square lattice and obtain the critical exponent $\nu=0.69\pm0.01$, which is close to the exact value $\nu=2/3$. As in the Ising case, we sample spin configurations with the Metropolis algorithm from a fixed ferromagnetic initial state.
The temperature runs from $T_{min}=0.91$ to $T_{max}=0.95$ with interval $\delta T=5\times 10^{-5}$.
We collect 20,000 configurations for each $T$.
The CNN architecture is the same as in the Ising case.
We scan $T_m$ between 0.91 and 0.95 with $\Delta T=5\times 10^{-5}$.
The resulting performance curves for system size $L=16$, 24, 32, 48, 64, and 128 and the finite-size scaling of the location of the minima are shown in Fig.~5(c) and Fig.~5(d).

In the two above studies we use the known results of the critical temperatures. In cases that critical temperatures are not known one can either introduce $T_c$ as an additional parameter in the finite-size scaling, or use a separate machine learning scheme, such as confusion,~\cite{Nieuwenburg17} to determine $T_c$.


\begin{thebibliography}{99}
\bibitem{y1y2}
Note that in the two-class case we find $y_2 \approx y_1$ in practice,
so the horizontal line $y_1 = y_2$ is close to $y_1 = 0$, but not exactly.
\end{thebibliography}
\end{document}